\documentclass[aps,preprint]{revtex4}%
\usepackage{amsfonts}
\usepackage{amsmath}
\usepackage{amssymb}
\usepackage{graphicx}
\usepackage{pdfpages}
\usepackage{float}
\usepackage{color}
\usepackage{xcolor}
\usepackage{caption}
\usepackage{lineno,hyperref}
\usepackage{changes}%
\hypersetup{colorlinks =true, allcolors = blue}
\providecommand{\U}[1]{\protect\rule{.1in}{.1in}}

\begin{document}
\begin{center}
{\Large Particle dynamics and shadow of a regular non-minimal magnetic black hole.
}

Ahmad Al-Badawi and M.Q. Owaidat

Department of Physics, Al-Hussein Bin Talal University, P. O. Box: 20,
71111, Ma'an, Jordan

\bigskip E-mail: ahmadbadawi@ahu.edu.jo, Owaidat@ahu.edu.jo

{\Large Abstract}
\end{center}

 In this paper, we study the dynamics of a test particle around a regular black hole (BH) in a non-minimal Einstein–Yang–Mills (EYM) theory and examine the BH shadow. The EYM theory is a non-minimally coupled theory in which curvature couples to non-Abelian gauge fields. We investigate particle motion with parameters in EYM BH for  massless and massive particles. This work provides the horizon structure, photon radius and inner stable circular orbit (ISCO) of a mass particle with EYM BH parameters. An analysis is provided of the effective potential as well as the possible orbits for test particles under various EYM BH parameters values. In timelike radial geodesics, we find that for smaller values of magnetic charge, particles following a timelike radial geodesic are more hasty in EYM BH, and hence arrive at the center faster than those traveling a Schwarzschild BH geodesic. However, at larger values of the magnetic charge, the inverse effect is observed. The effect of model parameters is investigated in order to study the ISCO, photon radius, orbit stability (Lyapunov exponent), and effective force on particles for the BH in the EYM theory. Finally, we investigate the BH shadow. We find that higher magnetic charge values and non-minimal coupling parameters result in smaller shadow radius values.  

\section{Introduction}
Gravitational theories coupled with electromagnetism have one specific characteristic: the Lagrangian include cross terms involving specific interactions, including scalar products of the Riemann tensor and its convolutions, and they are known as non-minimal theories. A non-minimal theory has numerous applications to cosmology and astrophysics. The non-minimal field theories coupled to gravity
five kinds are: first kind is scalar fields coupling with the spacetime curvature. The theories
were presented differently by different authors \cite{ey1,ey2,ey3,ey4,ey5}. Secondly, a non-minimal Einstein-Maxwell model of curvature-coupled
electromagnetic fields \cite{ey6,ey7,ey8,ey9}. Third kind is the Einstein-Yang-Mills (EYM) theories, or models with $SU(n)$
symmetry \cite{ey10,ey11}. The Einstein-Yang-Mills-Higgs model is examined in References \cite{ey12,ey13}. Last kind is the non-minimal Einstein-Maxwell-Axion models \cite{ey14}.

On the other hand, several aspects of star and BH physics have been discussed in
these types of non-minimally coupled theories. In particular, for any
event horizon,  equations of EYM with the gauge group SU(2) have infinite BH
solutions \cite{ey15}. Several studies on the EYM theory  and its BH
solutions have been conducted \cite{ey16,ey17,ey18,ey19}. 

In a paper by Balakin et al. \cite{ey20,ey21} presents a new exact spherically symmetric static solution of a non-minimal $SU(2)$ EYM theory with a cosmological and a Wu-Yang ansatz. The metric function shows explicitly a four-parameter family of exact solutions. These parameters are the non-minimal parameter, the cosmological constant, the magnetic charge, and the mass. A number
of studies have been presented to investigate the EYM BH, specifically the weak
and strong deflection gravitational lensings in \cite{ey22} and for 
rotating versions in \cite{ey23}. The rotating EYM BH shadow, the quasinormal modes, and the
quasiperiodic oscillations was studied in \cite{ey24}. \cite{ey25} investigated the
particle dynamics of an EYM BH with a cosmological constant. Effects of
thermal fluctuations on EYM BH with cosmological constant was examined \cite{ey26}. Very recently the spinorial wave equations and the greybody factors
in the background of EYM BH were investigated \cite{ey26b}. 
The purpose of this study is to explore the spacetime structure, particle
dynamics and BH shadow of a regular BH as described by the EYM theory coupled with
the $SU(2)$ gauge field \cite{ey20,ey21}. Because spacetime structure strongly influences the behaviour of geodesic
lines, test particle motion can also be used to test metric theories of
gravity. A wide range of references \cite{ey27,ey28,ey29,ey30,ey31,ey32,ey33,ey34,ey35,ey36,ey37,ey38,ey39,ey40,ey41,ey42,ey43,ey44,ey45,ey46,ey47,ey48,ey49,ey50,ey51,ey52,ey53} have examined the influence of spacetime curvature and gravity field parameters on particle dynamics. 

In the strong gravity regime, null geodesics form the image of a BH shadow. Photons with low angular momentum fall into the BH and form a dark area for a distant observer, whereas photons with high angular momentum will be deflected by the BH's gravitational potential. In contrast, photons of critical angular momentum orbit the BH indefinitely and surround the dark interior of the BH, forming the photon ring and the BH shadow. The first study of light deflection around a Schwarzschild BH was made by Synge \cite{sh1}, and then Luminet simulated a shadow photograph of the BH \cite{sh2}. Many researchers have been working on theoretical modelling of black hole shadows in recent years as a result of the detection of BH shadows \cite{sh3,sh4,sh5,sh6,sh7,sh8,sh9,sh10,sh11,sh12,sh13}.

Our study of particle dynamics and BH shadow in the background of a regular EYM BH is motivated by the following reasons: first, geodesic studies provide information about the
gravitational field around a BH. Moreover, examine whether the Schwarzschild BH and the EYM BH have distinct shadows. Another reason for this research is that
the EYM BH is a regular BH,   and regular BHs solutions (both
static and rotating), are thought to be one of the
solutions to the problem of the existence of singularities in General
Relativity \cite{sh67a,sh67b,sh67c,sh67d,sh67e,sh67f}. Furthermore, the spacetime under consideration is classified as
a non-minimal field theory. A non-minimal theory can provide exact solutions for wormholes, stars and BHs with electric and magnetic fields.
This is the outline of the paper: We briefly introduce the EYM BH spacetime in section 2 and write  equations of the geodesic  that describe the particle motion in the EYM BH spacetime. Section 3 goes over radial geodesics for null and timelike particles. Section 4 looked at circular timelike geodesics. We also investigate the Lyapunov exponent and the particle's force. In section 5 we study the shadow of the EYM BH. Finally, we conclude our findings and results.

\section{EYM BH spacetime and geodesic equations}

Balakin et al. \cite{ey21} present a new exact regular spherically symmetric solution of the non-minimal EYM theory with magnetic charge of the Wu-Yang gauge field. Taking into account their static spherically symmetric spacetime  with line element:
\begin{equation}
ds^{2}=-f\left( r\right) dt^{2}+\frac{dr^{2}}{f(r)}+r^{2}\left( d\theta ^{2}+\sin
^{2}\theta d\phi ^{2}\right)  \label{M1}
\end{equation}
where
\begin{equation}
f\left( r\right) =1+\left( \frac{r^{4}}{r^{4}+2\xi Q^{2}}\right) \left( 
\frac{Q^{2}}{r^{2}}-\frac{2M}{r}\right),
\end{equation}
where $M$ is the BH mass, $Q$ is the magnetic charge and $\xi$ represents the non-minimal parameter of the theory. 
When $Q=0$, metric given in Eq. (\ref{M1}) become  the
Schwarzschild (Schw) BH. The metric function $f(r)$ behaves as
\begin{equation}
\lim_{r\rightarrow 0}f(r)=1+\frac{r^{2}}{2\xi}-\frac{Mr^{3}}{\xi Q^{2}} +\mathcal{O}\left( r^{6}\right) .  \label{mf1}
\end{equation}
When approaching the source, Eq. (\ref{mf1}) reduces to 1, implying that EYM BH is regular. At finite positive r, the metric produces spacetime curvature singularities for $\xi <0$. Because there are no curvature singularities for $\xi >0$, we chose $\xi >0$. To justify this choice, we calculate the Ricci squared scalar:
\begin{eqnarray}
R_{\mu \nu }R^{\mu \nu } &=&\frac{\left( -120Q^{2}\xi
Mr^{5}-3Q^{2}r^{8}-144Q^{4}\xi ^{2}Mr-36Q^{2}\xi ^{2}+72Q^{4}\xi
r^{4}\right) ^{2}}{9\left( r^{4}+2Q^{2}\xi \right) ^{6}}  \nonumber \\
&&+\frac{\left( 3Q^{2}r^{4}+48Q^{8}\xi Mr-18Q^{4}\xi \right) ^{2}}{9\left(
r^{4}+2Q^{2}\xi \right) ^{4}}
\end{eqnarray}
\begin{equation}
\lim_{r\rightarrow 0}R_{\mu \nu }R^{\mu \nu }=\frac{9}{\xi ^{2}}.
\end{equation}
One may note that the BH is regular and for no curvature singularities  $\xi
>0$ must be held. 
\begin{figure}
{{\includegraphics[width=7.5cm]{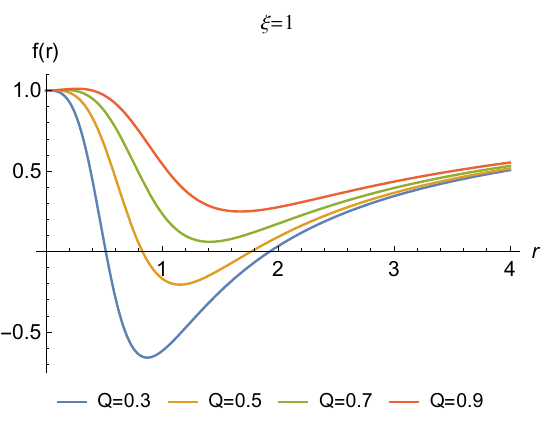} }}
    \caption{ Metric function of EYM BH at $\xi=1$ for different values of $Q$.} \label{fig1a}
\end{figure}
Figure \ref{fig1a} shows the metric function $f\left( r\right) $ as a function of radial
dependence for $\xi=1$, it shows two real roots which disappear as $Q$ increases. We note that, different values of $\xi$ lead to similar behavior for the metric function. 
Therefore, EYM BH has outer $\left( r_{+}\right) $ and inner $\left(
r_{-}\right) $\ horizons. To locate these event horizons one has to find the
roots of the metric function or
\begin{equation}
r^{4}+r^{2}\left( Q^{2}-2Mr\right) +2\xi Q^{2}=0.  \label{hor2}
\end{equation}
Because Eq. (\ref{hor2}) is a quartic equation, it has four roots; however, when we solve it, we get two real roots and two imaginary roots. The real positive roots of  Eq. (\ref{hor2}) give the locations of the horizons $r_{+}$ and $r_{-}<r_{+}$ as follows: 
\begin{equation}
r_{\pm }=\frac{M}{2}+\frac{1}{2}\sqrt{\left( M^{2}-\frac{2Q^{2}}{3}+\frac{A}{%
3\sqrt[3]{B}}+\frac{\sqrt[3]{B}}{3\sqrt[3]{2}}\right) }\pm  \label{hor1}
\end{equation}%
\begin{equation*}
\frac{1}{2}\sqrt{\left( 2M^{2}-\frac{4Q^{2}}{3}-\frac{A}{3\sqrt[3]{B}}-\frac{%
\sqrt[3]{B}}{3\sqrt[3]{2}}+\frac{8M\left( M^{2}-Q^{2}\right) }{4\left( \sqrt{%
\left( M^{2}-\frac{4Q^{2}}{3}+\frac{A}{3\sqrt[3]{B}}+\frac{\sqrt[3]{B}}{3%
\sqrt[3]{2}}\right) }\right) }\right) },
\end{equation*}%
where%
\begin{equation*}
A=\sqrt[3]{2}Q^{2}\left( Q^{2}+24\xi \right) ,
\end{equation*}%
\begin{equation*}
B=2Q^{6}+216M^{2}Q^{2}\xi -144Q^{2}\xi +2Q^{2}\sqrt{\left( Q^{4}+108M^{2}\xi
-72Q^{2}\xi \right) ^{2}-Q^{2}\left( Q^{2}+24\xi \right) ^{3}}.
\end{equation*}
\begin{figure}
    \centering
{{\includegraphics[width=7.5cm]{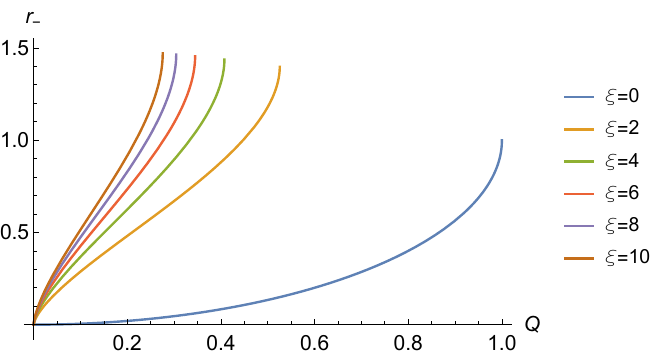} }}\qquad
{{\includegraphics[width=6 cm]{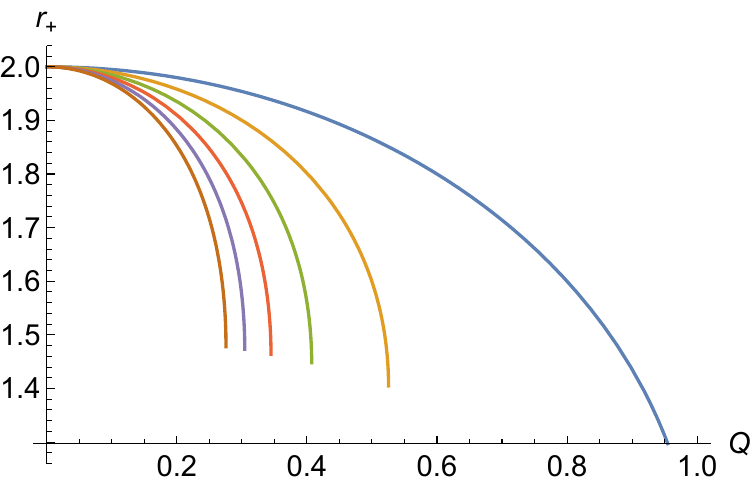}}}
    \caption{Inner (left) and outer (right) horizons for or various values of non minimal coupling parameter.} \label{fig1}
\end{figure}
\begin{figure}
    \centering
{{\includegraphics[width=7.5cm]{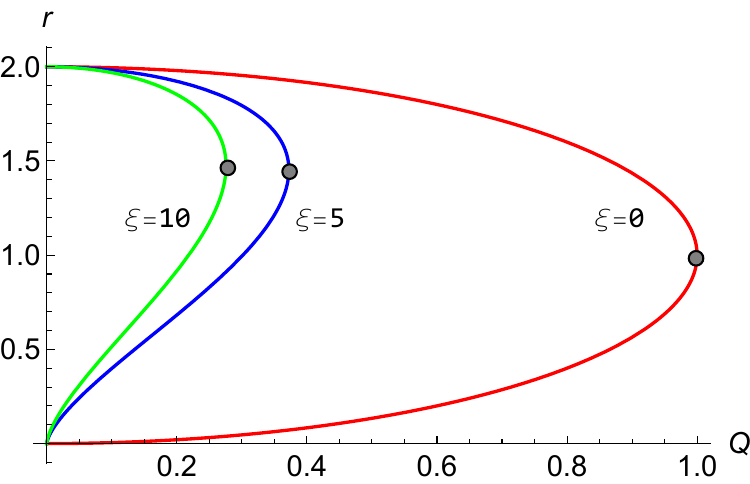} }}
    \caption{Inner and outer  horizons for  various values of $\xi$ and $Q$.} \label{fig2}
\end{figure}
Note that Eq. (\ref{hor1}) reduces to $r=2M$ in case $Q=0$. The parameters $\xi $ and $Q$ have an effect on the
event horizon as shown in Eq. (\ref{hor1}). Figure \ref{fig1} shows how the inner and
outer horizons of the regular EYM BH varies with the magnetic charge $Q$ for various
values of non-minimal coupling parameter $\xi$. According to the Fig.,
increasing the magnetic charge increases (decreases) its inner (outer) horizon radius. 
However, when $r_{+}=r_{-}$ the two horizons coincide, which means that we have an extremal BH. For a given $Q$ there is a critical value for $\xi_{cri}$ that makes
two event horizons reduce into one $r_{cri}$  satisfying the extremal conditions\begin{equation}    f(r_{cri})=0,\hspace{1cm}f'(r_{cri})=0
\end{equation} It is interesting to note that going beyond this critical value, $r>r_{cri}$ (e.g. when $\xi=1$ and $Q=0.7$ Eq. (\ref{hor2}) has no real roots) , one can see that event horizons no longer exist and the BH becomes a naked singularity.
Figure \ref{fig2} shows when the two horizons meet each others where the gray circles represent the location of extremal BH case. 

Now we  write equations of geodesic  for a massive particles around the
EYM BH. The Lagrangian for the metric given in Eq. (\ref{M1}) is 
\begin{equation}
2\mathit{L}=-f\left( r\right) {\dot{t}}^{2}+ \frac{{\dot{r}}%
^{2}}{f\left( r\right)}+r^{2}\left( {\dot{\theta}}^{2}+\sin ^{2}\theta {\dot{\phi}}^{2}\right) ,
\label{lag1}
\end{equation}%
where a dot indicates a derivative with respect to the affine
parameter $\sigma.$ The EYM BH have two Killing vectors, namely,
\begin{equation}
E=\frac{dL}{d\dot {t}}=-\left( 1+\frac{r^{2}\left(
Q^{2}-2Mr\right) }{r^{4}+2\xi Q^{2}}\right) \dot{t}=\text{constant}
\label{E1}
\end{equation}
\begin{equation}
\ell =\frac{dL}{d\dot {\phi}}=r^{2}\sin ^{2}\theta 
\dot {\phi }=\text{constant,}  \label{An1}
\end{equation}
where $E$ is the energy and $\ell$ is an angular momentum.
Choosing the initial
conditions $\theta =\pi /2$ and $\dot {\theta }=0$, the geodesic motion is confined to the equatorial plane. Using Eqs. (\ref{E1}) and (\ref{An1}), the geodesic equation becomes 
\begin{equation}
\left( \frac{dr}{d\phi }\right) ^{2}=\frac{r^{4}}{\ell ^{2}}\left[
E^{2}-\left( 1+\frac{r^{2}\left( Q^{2}-2Mr\right) }{r^{4}+2\xi Q^{2}}\right) %
\right] \left( \epsilon +\frac{\ell ^{2}}{r^{2}}\right) ,  \label{g1}
\end{equation}
where $\epsilon =0$ corresponds to null geodesics and $\epsilon =1$
corresponds to timelike geodesics. The $r$ equation as a
function of $t$ and $s$ can be obtained  as

\begin{equation}
\left( \frac{dr}{dt}\right) ^{2}=\frac{1}{E^{2}}\left[ E^{2}-\left( 1+\frac{%
r^{2}\left( Q^{2}-2Mr\right) }{r^{4}+2\xi Q^{2}}\right) \left( \epsilon +%
\frac{\ell ^{2}}{r^{2}}\right) \right] \left( 1+\frac{r^{2}\left(
Q^{2}-2Mr\right) }{r^{4}+2\xi Q^{2}}\right) ^{2}.  \label{g2}
\end{equation}%
\begin{equation}
\left( \frac{dr}{d\sigma}\right) ^{2}=E^{2}-\left( 1+\frac{r^{2}\left(
Q^{2}-2Mr\right) }{r^{4}+2\xi Q^{2}}\right) \left( \epsilon +\frac{\ell ^{2}%
}{r^{2}}\right) ,  \label{g3}
\end{equation}%
We may summarize the particle's motion using the three equations (\ref{g1}-\ref{g3}), just as we can for Schwarzschild BH. By Applying the
normalization condition $u^{\alpha }u_{\alpha }=-1,$ the equation
of motion can be written as
\begin{equation}
\left( \frac{dr}{d\sigma}\right) ^{2}+2V_{eff}\left( r\right) =E^{2},
\end{equation}%
where the effective potential, $V_{eff}$ is given by
\begin{equation}
2V_{eff}\left( r\right) =\left( 1+\frac{r^{2}\left( Q^{2}-2Mr\right) }{%
r^{4}+2\xi Q^{2}}\right) \left( \epsilon +\frac{\ell ^{2}}{r^{2}}\right) .
\label{veff1}
\end{equation}
The above equations are the primary equations governing radial and circular geodesic motion. We will now apply those equations in the next two sections.

\section{Radial timelike geodesics and particle trajectories}

Here, we examine the radial geodesics of particles in the EYM BH spacetime
described by Eq. (\ref{M1}). In this sense, $\overset{\cdot }{%
\phi }=0$  and $\ell =0$, Eq. (\ref{g3})
becomes%
\begin{equation}
\left( \frac{dr}{d\sigma}\right) ^{2}=E^{2}-\epsilon \left( 1+\frac{r^{2}\left(
Q^{2}-2Mr\right) }{r^{4}+2\xi Q^{2}}\right) .  \label{g8}
\end{equation}

\subsection{Null geodesics}

For radial null geodesics we choose $\epsilon =0$. Therefore Eq. (
\ref{g8}) becomes
\begin{equation}
\frac{dr}{d\sigma}=\pm E,  \label{g9}
\end{equation}%
where -(+) stands for ingoing (outgoing) photon. In
terms of $t,$ Eq. (\ref{g9}) reads 
\begin{equation}
\frac{dr}{dt}=\pm \left( 1+\frac{r^{2}\left( Q^{2}-2Mr\right) }{r^{4}+2\xi
Q^{2}}\right) .  \label{g10}
\end{equation}%
Integrating Eq. (\ref{g10}) we obtain%
\begin{equation}
\pm 2\left( t-t_{0}\right) =2r+\frac{Q^{3/2}}{2^{3/4}\xi ^{1/4}}\left[ \tan
^{-1}\left( \frac{-\xi ^{1/4}\sqrt{Q}+2^{1/4}r}{\xi ^{1/4}\sqrt{Q}}\right)
+\tan ^{-1}\left( \frac{\xi ^{1/4}\sqrt{Q}+2^{1/4}r}{\xi ^{1/4}\sqrt{Q}}%
\right) \right]  \label{time1}
\end{equation}%
\begin{equation*}
+\frac{Q^{3/2}}{2^{7/4}\xi ^{1/4}}\left[ \log \left( 2\sqrt{\xi }%
Q-2^{5/4}\xi ^{1/4}\sqrt{Q}r+\sqrt{2}r^{2}\right) -\log \left( 2\sqrt{\xi }%
Q+2^{5/4}\xi ^{1/4}\sqrt{Q}r+\sqrt{2}r^{2}\right) \right]
\end{equation*}%
\begin{equation*}
-M\log \left( 2\xi Q^{2}+r^{4}\right) .
\end{equation*}%

\begin{figure}
    \centering
{{\includegraphics[width=7.5cm]{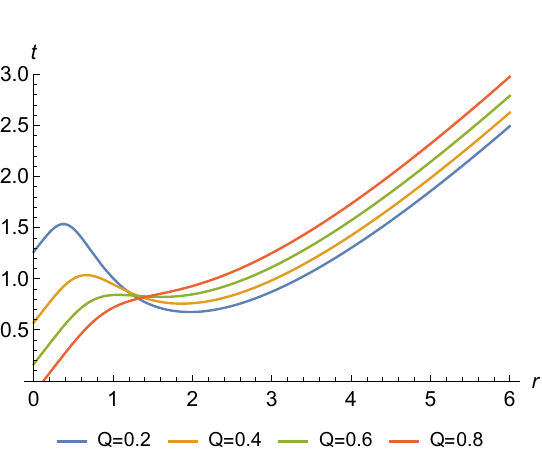} }}
    \caption{The relation of the coordinate time $t$  with the radius $r$ on a radial null geodesic.} \label{fig3}
\end{figure}

  In Fig. \ref{fig3} the time $t$ is plotted as a function of distance $r$ for different values of $Q$. It is seen  that the time  decreases as the BH magnetic charge
values increase. However, as the photon approaches the BH's horizon, a
turning point appears, after which the effect of the BH magnetic charge is
exactly the opposite on the time $t$. It indicates that for an external
observer the photon is moving faster toward the BH from the outside, and
after passing the turning point, it slows down. This observation
demonstrates that the photon in EYM\ BH travels to the horizon more quickly
than in the case of Schwarzschild BH.

\subsection{Timelike geodesics}

In this part, we study the equations of motion of a massive particle.
Taking $\epsilon =1$ and $l=0$, Eq. (\ref{g3}) becomes
\begin{equation}
\left( \frac{dr}{ds}\right) ^{2}=E^{2}-1-\frac{r^{2}\left( Q^{2}-2Mr\right) 
}{r^{4}+2\xi Q^{2}}.  \label{t123}
\end{equation}
By choosing $s$ as the proper time $\tau $, Eq. (\ref{t123})
becomes 
\begin{equation}
\frac{d^{2}r}{d\tau ^{2}} =\frac{2\xi Q^{2}r\left( Q^{2}-3Mr\right)
+r^{5}\left( Mr-Q^{2}\right) }{\left( r^{4}+2\xi Q^{2}\right) ^{2}}.
\end{equation}
The effective potential in case of radial timelike geodesics is given by 
\begin{equation}
V_{eff}\left( r\right) =\frac{1}{2}\left( 1+\frac{r^{2}\left(
Q^{2}-2Mr\right) }{r^{4}+2\xi Q^{2}}\right).
\label{veff2}
\end{equation}
The plot of Eq. (\ref{veff2}) is similar to Fig. \ref{fig1a}. Therefore, we observe that the effective potential curves are
concave up and become more stable as the value of magnetic charge decreases, 
thus stable orbits exist in EYM BH.

Let us now consider a particle initially at rest, or $\overset{\cdot }{r}=0$
, and that when it experiences gravitational attraction, it accelerates from
its initial location, $r=r_{0}$, towards the gravitating source. The relation between 
initial location and the constant of motion $E$ is given by%
\begin{equation}
E^{2}=1+\left( \frac{r_{0}^{4}}{r_{0}^{4}+2\xi Q^{2}}\right) \left( \frac{%
Q^{2}}{r_{0}^{2}}-\frac{2M}{r_{0}}\right) .
\end{equation}%
Changing the variable $r=r_{0}\cos ^{2}\eta /2$, $(0 \leq \eta \leq \pi)$ we obtain 
\begin{equation}
\left( \frac{dr}{d\tau }\right) ^{2}=Q^{2}\left[ \frac{r_{0}^{2}}{%
r_{0}^{4}+2\xi Q^{2}}-\frac{r^{2}}{r^{4}+2\xi Q^{2}}\right] +2M\left[ \frac{%
r^{3}}{r^{4}+2\xi Q^{2}}-\frac{r_{0}^{3}}{r_{0}^{4}+2\xi Q^{2}}\right] 
\label{enr1}
\end{equation}
When $Q=0,$ Eq. (\ref{enr1}), reduces to the corresponding equation in the
Schwarzschild BH. For in-falling particles the equation to be integrated is 
\begin{equation}
\frac{d\tau }{d\eta }=\frac{r_{0}\cos ^{2}\frac{\eta }{2}}{\sqrt{2M\left( 
\frac{r_{0}^{3}}{r_{0}^{4}+2\xi Q^{2}}\right) -Q^{2}\left( \frac{r_{0}^{2}}{%
r_{0}^{4}+2\xi Q^{2}}\right) -\frac{Q^{2}}{r_{0}^{2}\cos ^{2}\frac{\eta }{2}}%
}}.
\end{equation}
Integrating the above equation  $\left( \eta _{0}=0\text{ if }%
r=r_{0}\text{ and }\tau _{0}=0\right) $, we get 
\begin{equation}
\tau =\frac{2M}{R_{0}^{3/2}}\tan ^{-1}\left[ \frac{\sqrt{2R_{0}}\sin (\eta
/2)}{\sqrt{R_{0}\left( 1+\cos \eta \right) -2Q^{2}\left( \frac{r_{0}^{2}}{%
r_{0}^{4}+2\xi Q^{2}}\right) }}\right]   \label{f11}
\end{equation}%
\begin{equation*}
+\frac{\sqrt{R_{0}\left( 1+\cos \eta \right) -2Q^{2}\left( \frac{r_{0}^{2}}{%
r_{0}^{4}+2\xi Q^{2}}\right) }}{\sqrt{2}R_{0}}r_{0}\sin \left( \eta
/2\right) ,
\end{equation*}%
where 
\begin{equation}
R_{0}=2M\left( \frac{r_{0}^{3}}{r_{0}^{4}+2\xi Q^{2}}\right) -Q^{2}\left( 
\frac{r_{0}^{2}}{r_{0}^{4}+2\xi Q^{2}}\right) 
\end{equation}
\begin{figure}
    \centering
{{\includegraphics[width=7.5cm]{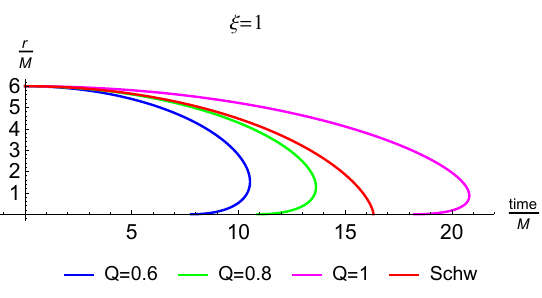} }}
    \caption{The variation of coordinate time $t$ and the proper time $\tau$ described by a test particle, starting at rest at $r=6M$ and falling  into the center.} \label{fig5}
\end{figure}
As shown in the equation above, when $Q=0$, we recover the Schwarzschild result namely, $\tau =\sqrt{\frac{%
r_{0}^{3}}{8M}}\left( \eta +\sin \eta \right)$. In Fig. \ref{fig5}, we plot Eq. (\ref{f11})  and compare it with the Schwarzschild case to
see the effect of magnetic charge on timelike radial motion. Figure \ref{fig5} clearly shows that increasing the value of the magnetic charge parameter $Q$ increases the proper time $\tau$. We find that for smaller values of $Q$, particles following a timelike radial geodesic are more hasty in EYM BH, and hence arrive at the center faster than those traveling a Schwarzschild BH geodesic. However, at larger values of the variable $Q$, the inverse effect is observed.

Let us now consider the motion of a massive particle when its trajectory is in
the radial direction of the BH. Using Eq. (\ref{g2}), we have
\begin{equation}
\left( \frac{dr}{dt}\right) ^{2}=\frac{1}{E^{2}}\left( 1+\frac{r^{2}\left(
Q^{2}-2Mr\right) }{r^{4}+2\xi Q^{2}}\right) ^{2}\left[ E^{2}-\left( 1+\frac{%
r^{2}\left( Q^{2}-2Mr\right) }{r^{4}+2\xi Q^{2}}\right) \right] .
\label{ert1}
\end{equation}
Integration of Eq. (\ref{ert1}) gives
\begin{equation}
\begin{split}
 \pm t=\frac{E}{\sqrt{E^{2}-1}}r+\frac{2^{5/4}\left( 2E^{2}-3\right) EQ^{3/2}%
}{16\left( E^{2}-1\right) \xi ^{1/4}}\left[ \tan ^{-1}\left( 1-\frac{2^{1/4}r
}{\xi ^{1/4}\sqrt{Q}}\right) -\tan ^{-1}\left( 1+\frac{2^{1/4}r}{\xi ^{1/4}
\sqrt{Q}}\right) \right] \\  +\frac{2^{1/4}\left( 2E^{2}-3\right) EQ^{3/2}}{16\left( E^{2}-1\right) \xi
^{1/4}}\left[ \log \left( 2\sqrt{\xi }Q-2^{5/4}\xi ^{1/4}\sqrt{Q}r+\sqrt{2}%
r^{2}\right) -\log \left( 2\sqrt{\xi }Q+2^{5/4}\xi ^{1/4}\sqrt{Q}r+\sqrt{2}%
r^{2}\right) \right] \\  \frac{2^{1/4}\left( 2E^{2}-3\right) EQ^{3/2}}{16\left( E^{2}-1\right) \xi
^{1/4}}\left[ 4M\sqrt{Q}\log \left( 2\xi Q^{2}+r^{4}\right) \right] .
\end{split}
\end{equation}
\begin{figure}
    \centering
{{\includegraphics[width=7.5cm]{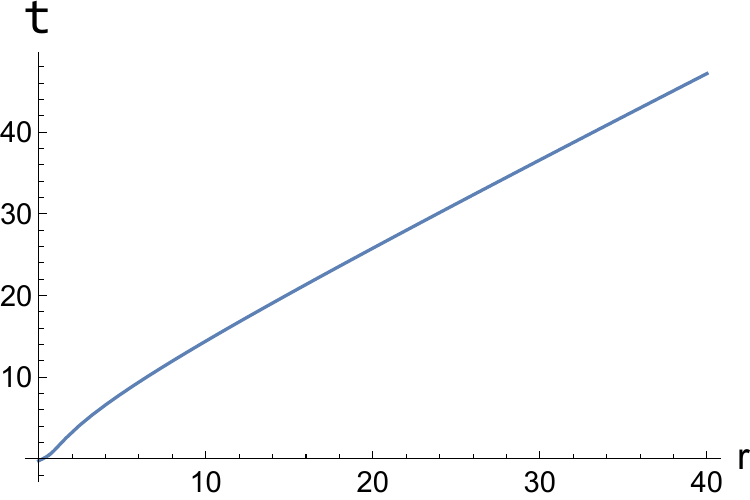} }}
    \caption{The relation of the coordinate time $t$  with the radial direction $r$. Here, $Q=0.4$,  $\xi=2$ and  $E=100$.} \label{fig6}
\end{figure}
Figure \ref{fig6} shows the relation between the coordinate time and the radial direction which is almost linear. 

\section{Circular timelike geodesics ($\epsilon =1$) and instability of Circular Orbit}

Here, we consider the circular motion of massive test particles
and photons for  $\ell \neq 0$. Based on Eq. (\ref{g3}), and with$f(r)=0, f^{\prime }(r)=0$ as conditions for circular orbits, the radial profiles for energy and angular momentum of circular orbits are given by
\begin{equation}
E^{2} =\frac{\left( 2\xi Q^{2}+r^{2}Q^{2}+r^{3}\left( r-2M\right) \right)
^{2}}{4\xi ^{2}Q^{4}+2\xi Q^{2}r^{3}\left( M+2r\right)
+2r^{6}Q^{2}+r^{7}\left( r-3M\right) },  \label{e22} 
\end{equation}
\begin{equation}
\ell ^{2}=\frac{2\xi Q^{2}r^{4}\left( Q^{2}-3Mr\right) +r^{8}\left(
Mr-Q^{2}\right) }{4\xi ^{2}Q^{4}+2\xi Q^{2}r^{3}\left( M+2r\right)
+2r^{6}Q^{2}+r^{7}\left( r-3M\right) }.  \label{an23}
\end{equation}%
\begin{figure}
    \centering
{{\includegraphics[width=7.5cm]{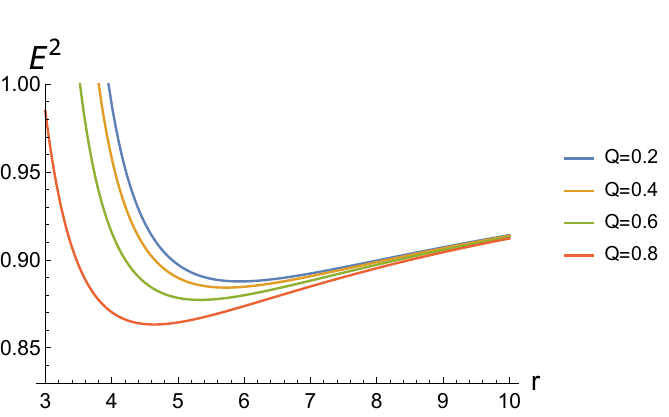} }}\qquad
{{\includegraphics[width=6 cm]{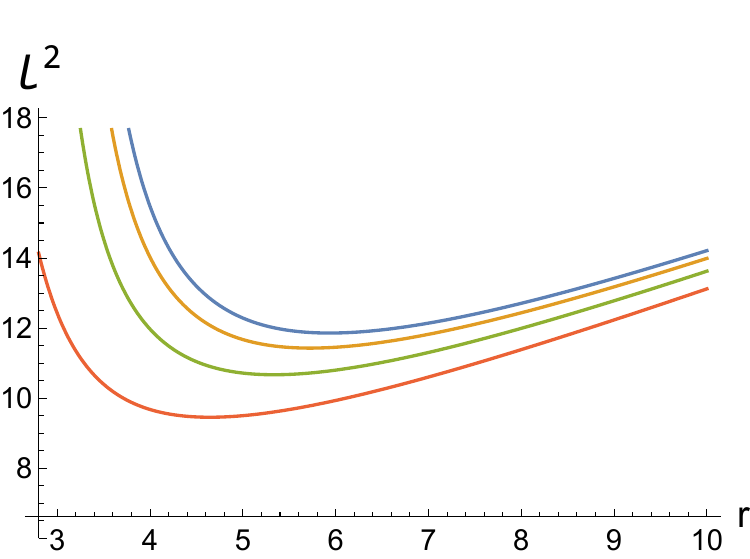}}}
    \caption{Radial dependence of energy (\ref{e22}) (left) and angular momentum (\ref{an23}) (right)  for different values of magnetic charge $Q$. Here, $\xi=1$.} \label{fig7}
\end{figure}
We see from Eq. (\ref{e22}) that for a physical acceptable motion the
constraint $4\xi ^{2}Q^{4}+2\xi Q^{2}r^{3}\left( M+2r\right)
+2r^{6}Q^{2}+r^{7}\left( r-3M\right) >0$ emerge naturally. Figure \ref{fig7} depicts
the radial dependence of $E^{2}$ and $\ell ^{2}$ for a test particle
moving in the equatorial plane on circular orbits. It is clear that as the
magnetic charge increases, result in a decrease in the minimum value of energy
and angular momentum. Hence, the circular orbits that correspond to constant values
of the test particle's energy and angular momentum move closer to the central
object. 

\subsection{Effective potential and the ISCO orbit}

\begin{figure}
    \centering
{{\includegraphics[width=7.5cm]{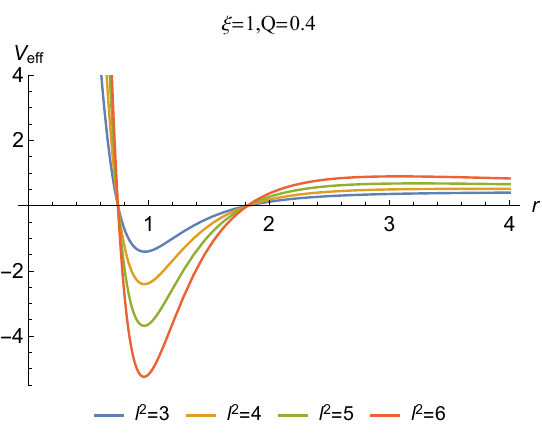} }}\qquad
    {{\includegraphics[width=7.5cm]{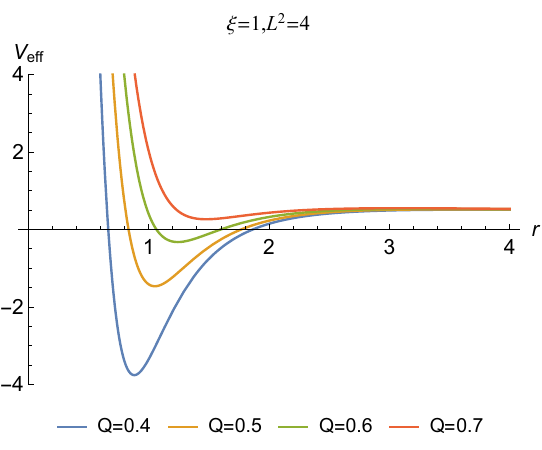}}}\qquad
    {{\includegraphics[width=7.5cm]{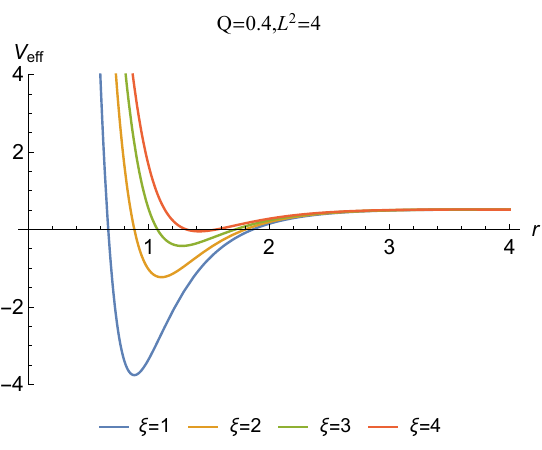}}}\caption{The figure shows $V_{eff}$ vs $r$ for the EYM BH for different values $\ell, Q$ and $\xi$.} \label{fig9}
\end{figure}

The effective potential in case of circular timelike geodesics is given by\begin{equation}
V_{eff}\left( r\right) =\frac{1}{2}\left( 1+\frac{r^{2}\left(
Q^{2}-2Mr\right) }{r^{4}+2\xi Q^{2}}\right) \left( 1+\frac{\ell ^{2}}{r^{2}}%
\right) .  \label{veff33}
\end{equation}
It is significant to remember that the angular momentum $\ell $
values play a key role in characterizing the potential orbits for time-like
geodesics. In other words, different choices of $\ell $ can result in a
variety of available orbits, based on changes in the effective potential's
shape. Finding the corresponding limiting values of $\ell $, that
characterize $V_{eff}\left( r\right) $, is therefore necessary. To elaborate this and to examine the stability of the equilibrium
circular motion of a massive test particle we  plot  $V_{eff}\left( r\right) $ given in Eq. (\ref
{veff33}). Figure \ref{fig9} represents the behavior of the effective potentials of EYM BH for different values of  $\ell,Q$ and $\xi$. One can see that as the
value of $\ell $ increases  $V_{eff}$ becomes more stable. The inverse effect is observed as the parameters $Q$ and $\xi$ are increased. There are however some parts of $V_{eff}$  that remain negative between the horizons, so the particles are bound between them. In general, for a given
energy larger than the asymptotic value of $V_{eff}$ all
curves are concave up and a particle can escape to infinity. There are bounded
orbits, which is a feature of $V_{eff}$ that is qualitatively equivalent to
its Newtonian counterpart.

\begin{figure}
    \centering
{{\includegraphics[width=7.5cm]{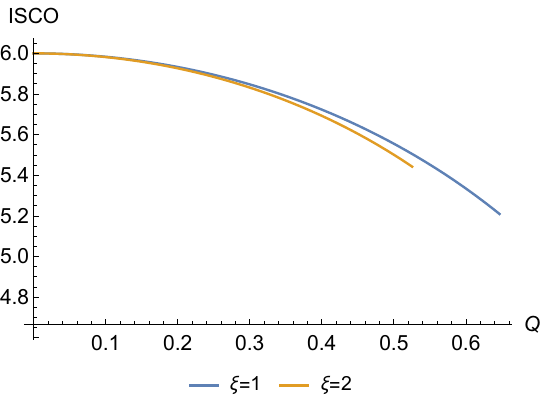} }}
    \caption{The parameters $\xi$ and $Q$ influence the ISCO radius.} \label{fig8}
\end{figure}

The ISCO orbit describes the minimal radius at which stable circular motion can still occur. To investigate the ISCO radius of the massive particle in EYM BH, we will use the standard conditions: 
\begin{equation}
V_{eff}=0,V_{eff}^{\prime }=0,V_{eff}^{\prime \prime }\geq 0,  \label{veff44}
\end{equation}%
where $V_{eff}$ is given by Eq. (\ref{veff33}).
 Based on Eq. (\ref{veff44}),
the following equation is obtained
\begin{equation}\begin{split}
\left( 2\xi Q^{2}+r^{2}Q^{2}+r^{3}\left( r-2M\right) \right) \Bigg( 4\xi
^{2}Q^{4}r\left( 4Q^{2}-15Mr\right)  +r^{7}\left(
-4Q^{4}+9MQ^{2}r+Mr^{2}\left( r-6M\right) \right) \\
+2\xi Q^{2}r^{4}\left(
-6M^{2}r-12Q^{2}r+M\left( Q^{2}+18r^{2}\right) \right) \Bigg) \geq 0
\end{split}
\label{rad11}    
\end{equation}
In the above equation, we obtain the radius of ISCO by solving for the radial coordinate. It is clear that Eq. (\ref{rad11}) is
extremely challenging to solve analytically. As a result, we present
numerical analysis with plots of the effects of both the magnetic charge and the non-minimal parameters on the ISCO radius of the massive particles. 
Solving Eq. (\ref{rad11}) numerically for $M=1$ and different values of $Q$ and $\xi$, the results for  the ISCO radius are  presented in Table \ref{tabl}.  
In Figure \ref{fig8}, the results tabulated in Table are reflected as well.
Figure \ref{fig8} shows how magnetic
charge effects ISCO radius of massive particles around regular non-minimal
EYM BH. It can be seen in the figure that the ISCO radius decreases as $Q$ increases.
\begin{table}[H]
    \centering
    \begin{tabular}{|c|c|c|c|}\hline
$\xi $ & $Q=0.3$ &  $Q=0.4$  &  $Q=0.5$  \\ \hline\hline
1 & 5.84731 & 5.72315 & 5.55531 \\ 
2 & 5.83159 & 5.69259 & 5.50098 \\ 
3 & 5.81561 & 5.66103 & 5.44319 \\ 
4 & 5.79938 & 5.62837 & 5.38131 \\ 
5 & 5.78287 & 5.59451 & 5.31452\\
\hline
    \end{tabular}
    \caption{Numerical results of the ISCO radius of the particles for various values of $\xi$ and $Q$ parameters when $M=1$.}
    \label{tabl}
\end{table}

\subsection{Lyapunov exponent for the unstable circular orbits}

\begin{figure}
    \centering
{{\includegraphics[width=7.5cm]{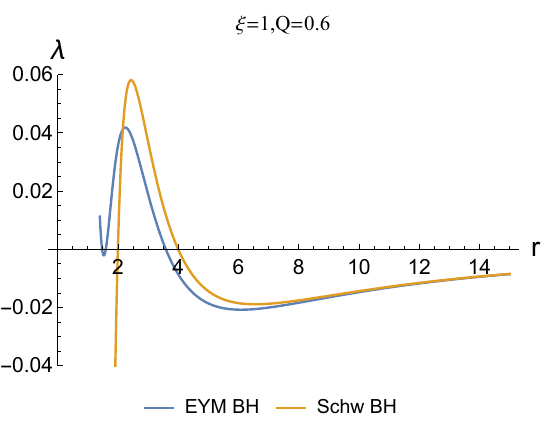} }}
    \caption{ Lyapunov exponent versus $r$ for the motion of
test particles around EYM BH.} \label{fig10}
\end{figure}

\begin{figure}
    \centering
{{\includegraphics[width=7.5cm]{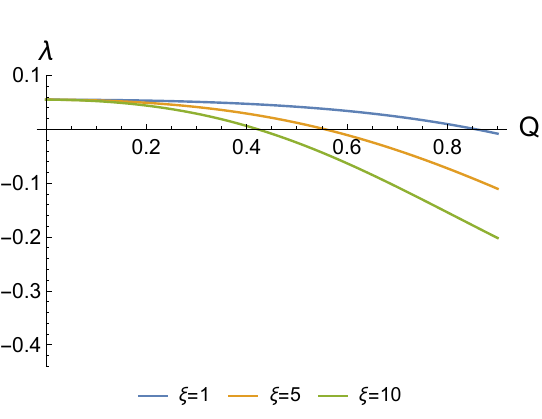} }}\qquad
    {{\includegraphics[width=7.5cm]{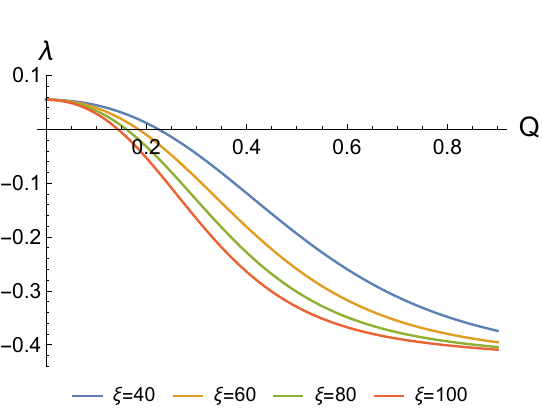}}}
    \caption{ Lyupanov exponent dependence for particles from $Q$ and small values of $\xi$ (left panel) and large values of $\xi$ (right panel).} \label{fig11}
\end{figure}

We investigate the stability (instability) of circular orbits for test particles orbiting around the EYM BH in this subsection. We consider the Lyapunov exponent as a measure of the average rate at which nearby trajectories can converge or diverge in phase space \cite{cardoso}. In general, a real Lyapunov exponent indicates a high sensitivity to the initial conditions between nearby trajectories. Additionally, the imaginary part of the quasinormal modes is governed by the Lyapunov exponent, which determines the timescale for instability at the equatorial null geodesic. The Lyapunov exponent is given by \cite{cardoso} 
\begin{equation}
\lambda =\frac{1}{\sqrt{2}}\sqrt{-\frac{V_{eff}^{\prime \prime }\left(
r_{c}\right) }{\left( \overset{\cdot }{t}\left( r_{c}\right) \right) ^{2}}}
\end{equation}%
\begin{equation}
    \begin{split}
    = \Bigg( -\frac{3}{r_{c}^{2}}-\frac{2r_{c}^{6}\left( Q^{2}-2Mr_{c}\right)
\left( 6\xi Q^{4}-4\xi Q^{2}r_{c}\left( 5M+4r_{c}\right) +r_{c}^{4}\left(
6Mr_{c}-5Q^{2}-8r_{c}^{2}\right) \right) }{\left( r_{c}^{4}+2\xi
Q^{2}\right) ^{4}}  \\
+\frac{r_{c}^{4}\left( 3Q^{2}-14Mr_{c}\right) -6\xi \left(
Q^{4}-2MQ^{2}r_{c}\right) }{\left( r_{c}^{4}+2\xi Q^{2}\right) ^{2}} \Bigg)
^{1/2},
    \end{split}
\end{equation}
where $r_{c}$ is the circular orbit obtained by solving the effective potential (\ref{veff33}) as $dV_{eff}/dr=0$. If $\lambda$ is complex, the orbits are stable, whilst for $\lambda$ that is real, the orbits are unstable.
 Figure \ref{fig10} shows the Lyapunov exponent around the EYM BH as a function of radial distance. According to Figure, the region where stable circular orbits can exist is minimized in the case of EYM BH as compared with Schwarzschild BH. However, Fig. \ref{fig11} shows that, the distance at which unstable orbits becomes stable for neutral particles shifts toward the central BH due to the increase of non-minimal parameters.  It can be seen that the value of $\lambda$ is real when $\xi=1$ (left panel of Fig. \ref{fig11}), which means that the orbits are stable when $\xi=1$. The values of $\lambda$ become negative as $\xi$ increases (right panel of Fig. \ref{fig11}), indicating unstable orbits. 

To find the critical exponent for instability of
the orbits we use \cite{ey5a,ey5b}
\begin{equation}
\gamma =\frac{T_{\lambda }}{T_{\tau }},
\end{equation}%
where, $T_{\lambda }$ is the instability time scale given by $T_{\lambda
}=1/\lambda $ and $T_{\tau }$ is the  period of a circular orbit which
can be determined from Eq. (\ref{An1}). Hence, 
\begin{equation}
    \begin{split}
    =\frac{\ell}{r^2sin^2\theta} \Bigg( -\frac{3}{r_{c}^{2}}-\frac{2r_{c}^{6}\left( Q^{2}-2Mr_{c}\right)
\left( 6\xi Q^{4}-4\xi Q^{2}r_{c}\left( 5M+4r_{c}\right) +r_{c}^{4}\left(
6Mr_{c}-5Q^{2}-8r_{c}^{2}\right) \right) }{\left( r_{c}^{4}+2\xi
Q^{2}\right) ^{4}}  \\
+\frac{r_{c}^{4}\left( 3Q^{2}-14Mr_{c}\right) -6\xi \left(
Q^{4}-2MQ^{2}r_{c}\right) }{\left( r_{c}^{4}+2\xi Q^{2}\right) ^{2}} \Bigg)
^{-1/2}
    \end{split}
\end{equation} 
The critical exponent $\gamma$ reveals whether gravitational wave signals can be detected. To detect gravitational signals generated by perturbation, therefore, the following requirements must be met: $T_{\lambda }<T_{\tau}$.
\subsection{Effective force}

\begin{figure}
    \centering
{{\includegraphics[width=7.5cm]{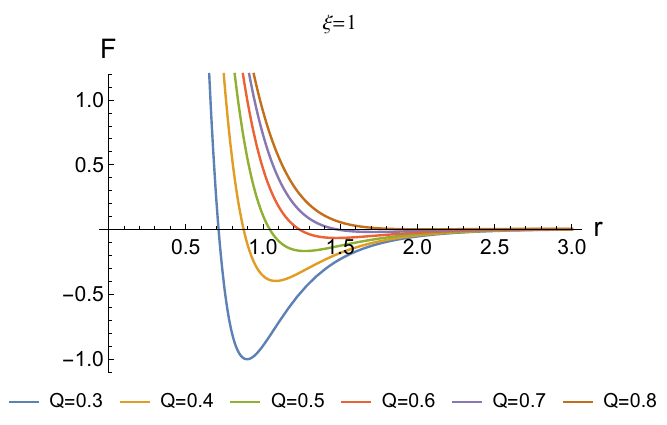} }}
    \caption{ Effective force versus the $Q$ and $\xi$ parameters.} \label{fig12}
\end{figure}

An effective force acts on particles to determine whether or not they are attracted to or moving away from the BH. The effective force on the particle can be calculated as  
\begin{equation*}
F=-\frac{1}{2}\frac{dV_{eff}\left( r\right) }{dr}
\end{equation*} 
\begin{equation}
=\frac{2M\xi Q^{2}}{(2\xi Q^{2}+r^{4})^{2}}+\frac{2M\xi^{2} Q^{4}}{r^{3}(4\xi Q^{2}+r^{4})^{2}}+\frac{4\xi Q^{2}r}{(2\xi Q^{2}+r^{4})^{2}}+\frac{2 Q^{2}r^{3}}{(2\xi Q^{2}+r^{4})^{2}}-\frac{3Mr^{4}}{(2\xi Q^{2}+r^{4})^{2}}-\frac{r^{5}}{(2\xi Q^{2}+r^{4})^{2}}.
\label{for1}
\end{equation}
The first four terms are repulsive force due to non-minimal EYM theory coupling, while the last two terms are attractive. Nevertheless, we plot Eq. (\ref{for1}) in Fig. \ref{fig12} to study the effect of $Q$ and $\xi$ on all terms of the force.  It is important to investigate the effective force because the point at which the force is zero, i.e. $r = r_{c}$, it gives the location of the stationary point corresponding to circular geodesics. According to Fig. \ref{fig12}, a massive particle has stable circular orbits and its force is negative for small values of Q, indicating that the force is attractive. For larger values, however, the force is positive, implying that the force is repulsive. In general, when the force is negative, photons are attracted to the BH and are pulled back towards it. It is interesting to note that for $r < r_{c}$ the force is attractive. However, for $r_{c} < r$  the force tends to zero and photons experience no force. 

\section{EYM BH shadow}

A BH shadow's contour can be found by separating the null geodesic
equations in the spacetime metric (\ref{M1}), using the Hamilton-Jacobi equation:%
\begin{equation}
\frac{\partial \mathcal{S}}{\partial \sigma }=-\frac{1}{2}g^{\mu \nu }\frac{%
\partial \mathcal{S}}{\partial x^{\mu }}\frac{\partial \mathcal{S}}{\partial
x^{\nu }},
\end{equation}%
where $\mathcal{S}$ is the Jacobi action. Consider the following Jacobi
action separable solution%
\begin{equation}
\mathcal{S}=-Et+\ell \phi +\mathcal{S}_{r}\left( r\right) +\mathcal{S}%
_{\theta }\left( \theta \right) ,
\end{equation}%
where $E$ and $\ell $ are defined in Eqs. (\ref{E1}) and (\ref{An1}). We can obtain the equations of
motion namely, 
\begin{equation}
\frac{dt}{d\sigma }=\frac{E}{f}, 
\end{equation}
\begin{equation}
\frac{d\phi }{d\sigma }=-\frac{\ell }{
r^{2}\sin ^{2}\theta },
\end{equation}
\begin{equation}
r^{2}\frac{dr}{d\sigma }=\pm \sqrt{\mathcal{R}\left( r\right) },  \label{R1}
\end{equation}
\begin{equation}
 r^{2}\frac{d\theta }{d\sigma }=\pm \sqrt{\Theta \left( \theta \right) } ,  
\end{equation}
where 
\begin{equation}
\mathcal{R}\left( r\right) =r^{4}E^{2}-\left( \mathcal{K}+\ell ^{2}\right)
r^{2}f,  \label{R2}
\end{equation}
\begin{equation}
\Theta \left( \theta \right) =\mathcal{K}-\ell ^{2}\cot \theta ,
\end{equation}
and $\mathcal{K}$ is the Carter separation constant.
Let us define the two quantities $\eta =\frac{\mathcal{K}}{E^{2}}$ and $%
\zeta =$ $\frac{\ell }{E}$ (stands for the impact parameters). In general,
it is well known that shadow casts can be obtained by using unstable null
circular orbits. Therefore, Eq. (\ref{R1}) can be rewritten as 
\begin{equation}
\left( \frac{dr}{d\sigma }\right) ^{2}+V_{eff}=0,
\end{equation}%
where%
\begin{equation}
V_{eff}\left( r\right) =E^{2}\left[ \frac{\eta +\zeta ^{2}}{r^{2}}\left( 1+%
\frac{r^{2}\left( Q^{2}-2Mr\right) }{r^{4}+2\xi Q^{2}}\right) -1\right] .
\end{equation}%
Circular orbits correspond to the maximum effective potential, and the
unstable photons should satisfy the following conditions:%
\begin{equation}
V_{eff}\left( r\right) \left\vert _{r=r_{ph}}\right. =0,V_{eff}^{\prime
}\left( r\right) \left\vert _{r=r_{ph}}\right. =0, \label{v33}
\end{equation}%
or alternatively,%
\begin{equation}
\mathcal{R}\left( r\right) \left\vert _{r=r_{ph}}\right. =0,\mathcal{R}%
_{eff}^{\prime }\left( r\right) \left\vert _{r=r_{ph}}\right. =0.  \label{R3}
\end{equation}%
where $r_{ph}$ is the radius of the unstable photon orbit. We can obtain $r_{ph}$ using Eqs. (\ref{v33}) and (\ref{R3}) which is nothing but the solution of \begin{equation}
r_{ph}f'(r_{ph})=2f(r_{ph}),    
\end{equation}or \begin{equation}
    4\xi ^{2}Q^{4}+2\xi Q^{2}r^{3}\left( M+2r\right)
+2r^{6}Q^{2}+r^{7}\left( r-3M\right)=0 \label{rc12} 
\end{equation}
Equation (\ref{R3}
) yields the following result
\begin{equation}
\eta +\zeta ^{2}=\frac{r_{ph}\left( r_{ph}+2\xi Q^{2}\right) }{r_{ph}+2\xi
Q^{2}+r_{ph}\left( Q^{2}-2Mr_{ph}\right) }.
\end{equation}

\begin{figure}
    \centering
{{\includegraphics[width=7.5cm]{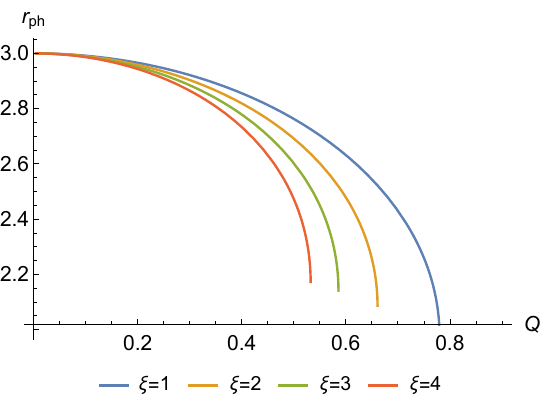} }}
    \caption{Dependence of the photon radius from the magnetic charge and the non-minimal coupling parameters.}\label{fig20}
\end{figure}

\begin{figure}
    \centering
{{\includegraphics[width=7.5cm]{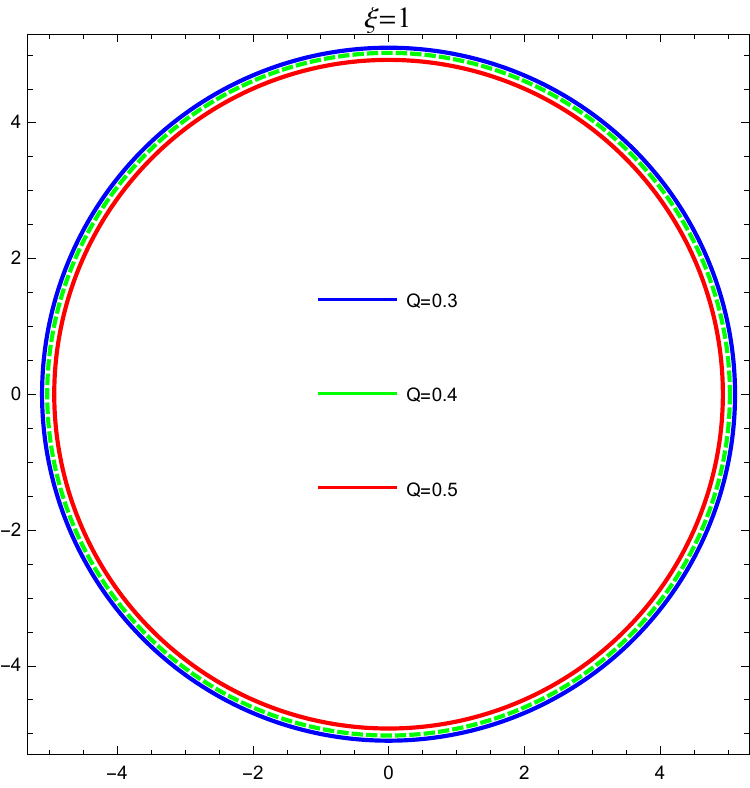} }}\qquad
    {{\includegraphics[width=7.5cm]{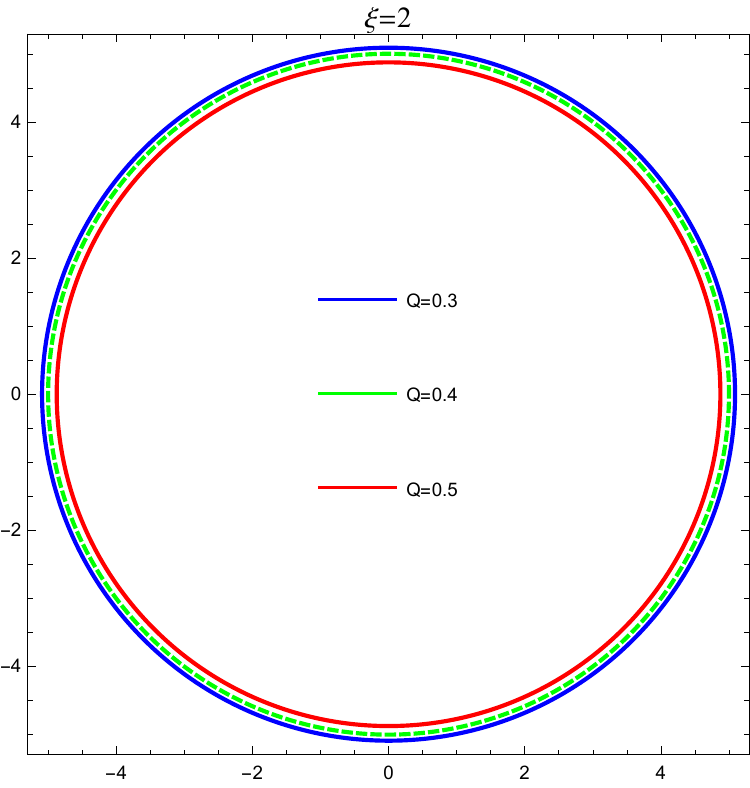}}}\qquad
    {{\includegraphics[width=7.5cm]{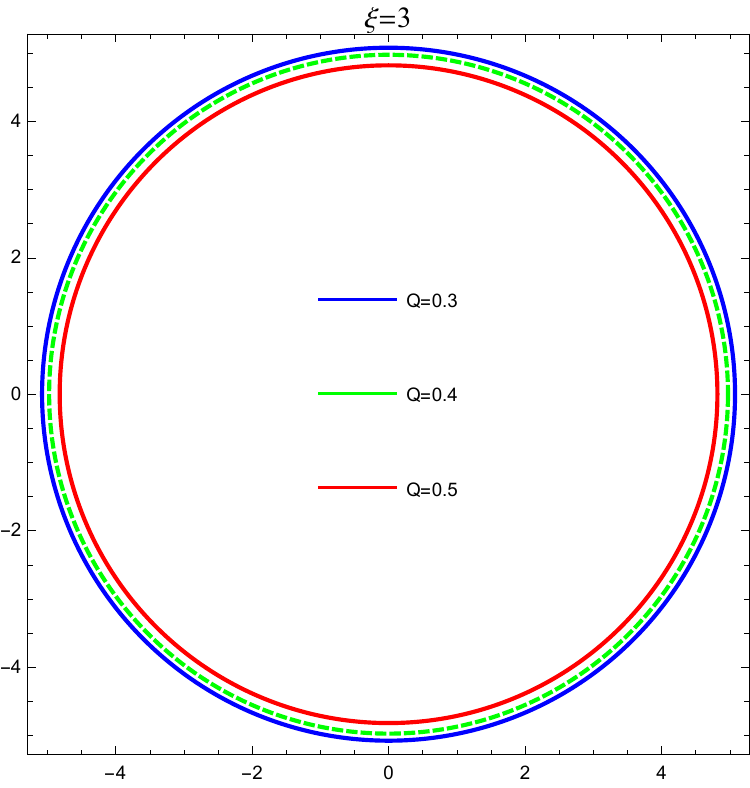}}}\qquad
    {{\includegraphics[width=7.5cm]{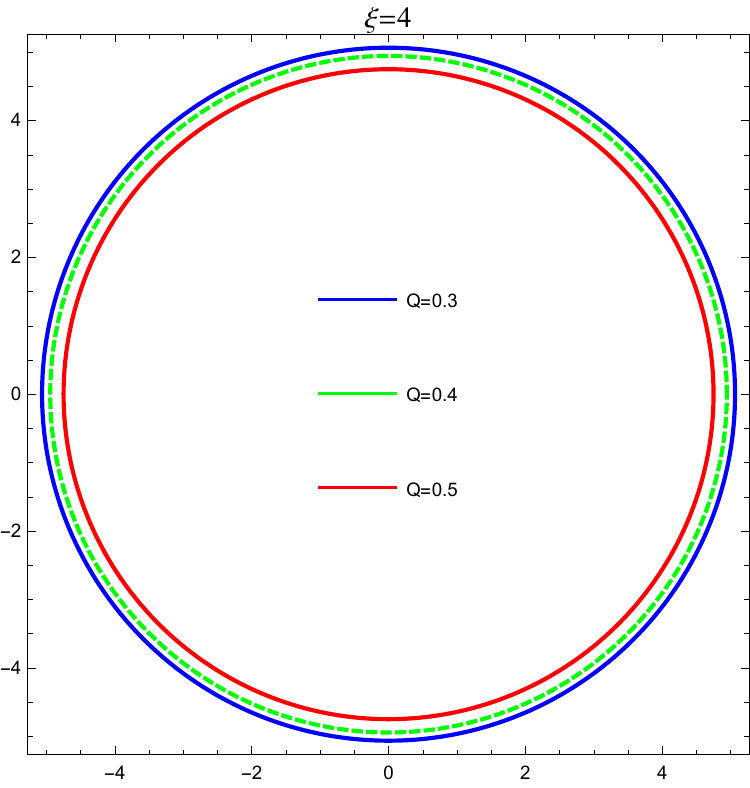}}}
    \caption{ Shadows of EYM BH for different values of magnetic charge and non-minimal coupling  parameters.}
    \label{fig10a}
\end{figure}

\begin{figure}
    \centering
{{\includegraphics[width=7.5cm]{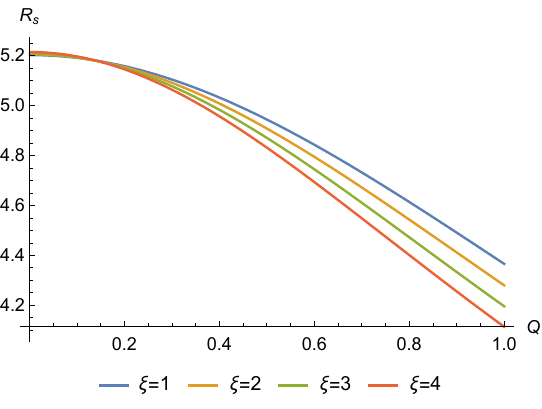} }}\qquad
    {{\includegraphics[width=7.5cm]{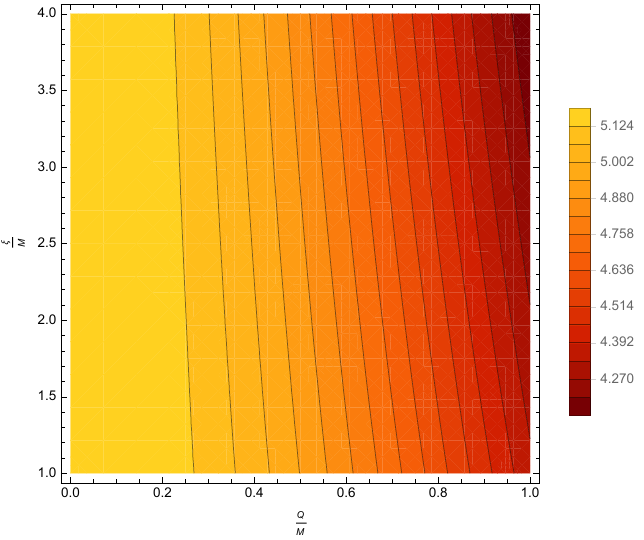}}}
    \caption{Variation of the shadow observable $R_{s}$ (left) and the contour maps for EYM BH shadow observable $R_{s}$ ( (right).} \label{f15}
\end{figure}

Analytically, Eq. (\ref{rc12}) is extremely difficult to solve. Due to this, we present numerical analysis and plots showing the effects of both the magnetic charge and non-minimal parameters on the mass particles' photon (circular) radius and the radius of the BH shadow. The results of
the photon radius $r_{ph}$ and the radius of the BH shadow $R_{s}$ are presented in Table \ref{tab2}. Our findings show that for a fixed value of non-minimal parameter $\xi$, the EYM BH shadow decreases as the magnetic charge increases. Furthermore, as value of the  parameter $\xi$ increases, the radius of the shadow decreases. Figure \ref{fig20} shows how magnetic charge effects the  $r_{ph}$ of massive particles around a regular
non-minimal EYM BH. One can see that, $r_{ph}$  decreases with the increase of $Q$ and $\xi$. One also note that, the magnetic charge shifts the $r_{ph}$
 towards the horizon of BH as it increases.
  \begin{table}[H]
     \centering
     \begin{tabular}{|c|c|c|c|c|c|c|c|c|} \hline
        & \multicolumn{2}{c}{$\xi =1$} &  \multicolumn{2}{|c|}{$\xi =2$} &  \multicolumn{2}{|c|}{$\xi =3$} &  \multicolumn{2}{|c|}{$\xi =4$}  \\ \hline 
 &$r_{ph}/M$ & $R_{s}/M$ & $r_{ph}/M$ & $R_{s}/M$ & $r_{ph}/M$ & $R_{s}/M$ & 
$r_{ph}/M$ & $R_{s}/M$  \\ \hline
$Q=0.3$ & $2.92146$ & $5.10407$ & $2.90345$ & $5.09128$ & $2.88463$ & $
5.07774$ & $2.86492$ & $5.06379$  \\ 
$Q=0.4$ & $2.85557$ & $5.02841$ & $2.81891$ & $5.0023$ & $2.77849$ & $4.97436
$ & $2.73316$ & $4.94417$  \\ 
$Q=0.5$ & $2.76282$ & $4.92471$ & $2.69165$ & $4.8761$ & $2.60205$ & $4.81958
$ & $2.47258$ & $4.74946$  \\ \hline
     \end{tabular}
     \caption{Numerical results for various parameters of the EYM BH.  The parameters are the photon radius $r_{ph}$ and the radius of the BH shadow $R_{s}$ (impact parameter). The 
case $Q=0$ corresponds to the Schw BH $r_{ph}/M=3$ and $R_{s}/M =5.19615$.}
     \label{tab2}
 \end{table} It has been reported \cite{ey60} that the real shadow of the BH seen on an observer's frame is described by celestial coordinates. Therefore, let us define the celestial coordinates  $X$ and $Y$ by \begin{equation}
X=\lim_{r_{0}\rightarrow \infty }\left( -r_{0}\sin \theta _{0}\left. \frac{
d\phi }{dr}\right\vert _{r_{0},\theta _{0}}\right) ,  \label{x11}
\end{equation}
\begin{equation}
Y=\lim_{r_{0}\rightarrow \infty }\left( r_{0}\left. \frac{d\theta }{dr}
\right\vert _{r_{0},\theta _{0}}\right) ,  \label{y11}
\end{equation}
 where $(r_{0},\theta_{0})$ are the position coordinates of the observer. Assuming the observer is on the equatorial hyperplane, Eqs. 
 (\ref{x11}) and (\ref{y11}) follow \begin{equation}
X^{2}+Y^{2}=R_{s}^{2}=\eta +\zeta ^{2}.  \label{xy1}
\end{equation} 
Figure \ref{fig10a} depicts the shadow cast by the EYM BH under the influence of magnetic charge and non-minimal parameters. According to Fig. \ref{fig10a}, increasing magnetic charge causes BH shadow to diminish. Further, the radius of BH shadow decreases with an increase in the non-minimal coupling parameter. This is because the larger values of $\xi$ weakens the strength of gravity so that the instability area around EYM BH decreases and thus the photon radius takes smaller values. 
The presence of parameters $Q$ and $\xi$ seems to affect the approximate size of the shadow in a significant way. Further, to show how shadow size varies with $(Q, \xi)$  we plot variation of the shadow observable $R_{s}$ and the contours map for the EYM BH shadow observable of $R_{s}$ in the $ Q, \xi$ parameter space in Fig. \ref{f15} \cite{g12}. It is evident that the shadow radius $R_{s}$ of EYM BH decreases with both $Q$ and $\xi$ increasing.

\section{Conclusion}

    This work examined the dynamics of neutral particles following the Lagrangian formalism, and the BH shadow of a spherically symmetric regular EYM solution. We found out that the effective potential of the neutral particle moving in radial timelike geodesics  in the spacetime of the regular EYM BH decreases with increasing values of the magnetic charge. According to our study, particles following a timelike radial geodesic in EYM BH are more hasty and thus reach the center faster than those following a Schwarzschild BH geodesic for lower values of $Q$. Moreover, we demonstrated that as the magnetic charge  of the EYM BH increases, causes a decrease in the minimum value of energy and angular momentum. Consequently, the circular orbits that correspond to constant values of the test particle's energy and momentum move closer to the central object. We have analyzed numerically the influence of of both the magnetic charge and the non-minimal parameters on the ISCO radius of the massive particles.  It was  found that, the ISCO radius decreases when the magnetic charge and the non-minimal parameters increase. We also presented numerically the radius of stable circular orbit and found that it decreases as both $Q$ and $\xi$ increase.  In our study of the instability of orbits of test particles, we found that increasing non-minimal parameter shifts the distance at which unstable orbits become stable toward central BH. We also found that, the force becomes more attractive as $Q$ decreases and does not change much with radial distance as it increases. We also conclude that non-minimal theory coupling the gravitational field with other fields by using the curvature tensor cross terms produces repellent forces, hence particles move away from BH.
 Finally, we studied the BH's shadow. It is found that the BH shadow radius monotonically decreases with both $Q$ and $\xi$ increased. Our findings show that both magnetic charge and the non-minimal parameter have an effect on the shadow of BH. The obtained result demonstrates that the shadow radius in EYM BH is less than the Schwarzschild radius and decreases monotonically as $Q$ and $\xi$ increase. As a result, the presence of the magnetic charge Q reduces the size of the shadow. 
\\ \\
{\LARGE Acknowledgements}\newline
We are thankful to the Editor and anonymous Referees for their constructive suggestions and comments.

{\Large Declarations}
\\
Conflict of interest: We have no funding, associated data, and conflict of interest/competing interests 
\newline

\end{document}